\documentclass[%
reprint,
superscriptaddress,
amsmath,amssymb,
aps,
]{revtex4-2}

\usepackage{graphicx}
\usepackage{dcolumn}
\usepackage{bm}
\usepackage{epstopdf}
\usepackage{booktabs}
\usepackage{tabularx}
\usepackage{makecell}
\usepackage{amsmath}
\usepackage [linesnumbered,lined,ruled,commentsnumbered]{algorithm2e}
\usepackage{subfigure}  
\usepackage{float}  
\usepackage{amssymb}
\usepackage{multirow}
\usepackage{colortbl}
\usepackage{color}
\usepackage{array}
\usepackage{soul}
\definecolor{darkblue}{rgb}{0,0,1}
\definecolor{darkred}{rgb}{0.75,0,0}

\newcolumntype{C}[1]{>{\centering\arraybackslash}m{#1}}
\newcolumntype{L}[1]{>{\raggedright\arraybackslash}m{#1}}
\newcolumntype{T}[1]{>{\centering\arraybackslash}t{#1}}
\definecolor{Lavender}{RGB}{230,230,250}

\begin{document}
	
	\preprint{APS/123-QED}
	
	\title{Detecting the driver nodes of temporal networks}
	
\author{Tingting Qin}
\affiliation{Center for Systems and Control, College of Engineering, Peking University, Beijing 100871, China}
\author{Gaopeng Duan}
\affiliation{Center for Systems and Control, College of Engineering, Peking University, Beijing 100871, China}
\author{Aming Li}
\thanks{Corresponding author: amingli@pku.edu.cn}
\affiliation{Center for Systems and Control, College of Engineering, Peking University, Beijing 100871, China}
\affiliation{Center for Multi-Agent Research, Institute for Artificial Intelligence, Peking University, Beijing 100871, China}
	
	
	\begin{abstract}
		Detecting the driver nodes of complex networks has garnered significant attention recently to control complex systems to desired behaviors, where nodes represent system components and edges encode their interactions.
		Driver nodes, which are directly controlled by external inputs, play a crucial role in controlling all network nodes. While many approaches have been proposed to identify driver nodes of static networks, we still lack an effective algorithm to control ubiquitous temporal networks, where network structures evolve over time.
		Here we propose an effective online time-accelerated heuristic algorithm (OTaHa) to detect driver nodes of temporal networks.
		Together with theoretical analysis and numerical simulations on synthetic and empirical temporal networks, we show that OTaHa offers multiple sets of driver nodes, and noticeably outperforms existing methods in terms of accuracy and execution time.
		We further report that most edges are redundant in controlling temporal networks although the complete instantaneous signal-carrying edges cannot be guaranteed.
		Moreover, removing edges with high edge betweenness (the number of all-pairs shortest paths passing through the edge) significantly impedes the overall controllability.
		Our work provides an effective algorithm and paves the way for subsequent explorations on achieving the ultimate control of temporal networks.
	\end{abstract}
	
	\maketitle
	
	
	\section{Introduction}


Complex networks, consisting of nodes representing components within a system and edges indicating their connections or relationships, serve as a highly effective framework for modeling a wide variety of natural and engineered systems~\cite{carrington2005models,stein2010complex,wang2023stability,jalili2017information,souma2003complex,angulo2016qualitative}.
In the field of network science, a significant amount of efforts have been invested in understanding, predicting and controlling the behavior of complex systems modeled as static networks, where links or connection patterns are fixed between nodes.
However, in many application cases, such as email communication, transportation and social interactions, the empirical networks are far from static with nodes and edges appearing and disappearing over time~\cite{masuda2016guide}.
It is increasingly recognized that real-world systems display intricate dynamical patterns not captured by static networks.
Indeed, some important results have been reported regarding temporal networks~\cite{estrada2022communicability,holme2012temporal,zhang2021higher,rocha2014random} (also known as time-varying networks, evolving networks, or dynamic networks), where the structure, as well as the attributes associated with nodes and edges, change over time.

By combining complex networks and control theory, recent progress uncovers the advantages of temporal networks on influencing or controlling complex systems to achieve desired behavior~\cite{li2017fundamental}.
Specifically, controllability is a key notion in the control theory, characterizing the ability of a dynamical system that can be driven from any initial state to any desired final state in finite time by the appropriate choice of external inputs~\cite{meng2023control,duan2019energy,duan2019energyACS}. However, in large-scale systems, it is challenging to determine the controllability by well-developed theoretical frameworks due to the fact that parameters of realistic systems are often measured with some non-negligible noise. 
This spurs researchers to focus on the interplay between the network structure and the corresponding controllability.
Lin studied the concept of structural controllability for linear time-invariant control systems~\cite{lin1974structural}.
By using the maximum matching algorithm, Liu~\textit{et al.}~proposed an efficient method to detect the minimum number of driver nodes for arbitrary directed networks~\cite{liu2011controllability}, which has stimulated further interest in exploring the structural controllability of complex networks~\cite{liu2016control,bassett2017network,albert2000error,gao2014target,jia2013emergence}.
P{\'o}sfai~\textit{et al.}~recently extended the concept of structural controllability of static networks to temporal networks and utilized the max-flow algorithm to determine the maximum controllable subspace formed by all controllable nodes of the set of driver nodes~\cite{posfai2014structural}.

Despite these important achievements, the problem of finding the minimum number and all corresponding sets of driver nodes for given temporal networks remains an NP-hard problem. This is due to the exponential increase in configuration choices to be tested with respect to the number of nodes in the network.
In recent years, several algorithms have been proposed to address the problem.
Ravandi~\textit{et al.}~developed a heuristic algorithm to yield multiple sets of driver nodes~\cite{ravandi2019identifying} and Srighakollapu~\textit{et al.}~presented a greedy algorithm to obtain an approximately optimal solution~\cite{srighakollapu2021optimizing}.
These algorithms have shown promising results and important applications in some systems.
However, we still lack an efficient algorithm that can improve the quality of identified driver node sets (namely, keeping the number of identified driver nodes as consistent as possible with the optimal solution) while reducing time complexity and computational cost.

Here, we present an online time-accelerated heuristic algorithm (OTaHa) to effectively detect driver nodes in temporal networks.
Our proposed algorithm not only provides a highly accurate approximate solution but also achieves a substantial reduction in execution time.
By theoretical analysis, we show that OTaHa can provide an upper bound of the number of driver nodes.
In numerical simulations, it is shown that our proposed algorithm reduces the execution time by a factor of 3 to 500 in comparison to the previous greedy algorithm.
Specifically, our approach involves mapping a set of driver nodes to the dimension of its corresponding maximum controllable subspace.
And the submodularity of the set function allows for a reduction in the number of times that the marginal gains must be updated during each iteration. 
Subsequently, by delving into the update process, the modified residual graph is presented.
A significant breakthrough is achieved by converting the max-flow in the modified residual graph to the marginal gain.
Beyond theoretical analyses, we demonstrate the effectiveness of our algorithm over a series of synthetic and empirical temporal networks, and present its possible applications in exploring the controllability of temporal networks, such as the role of edges and the evolution of controllability in the context of edge attacks.

%

\section{Model\label{preliminaries}}
\begin{figure*}[htbp]
	\centering
	\includegraphics[width=.95\textwidth]{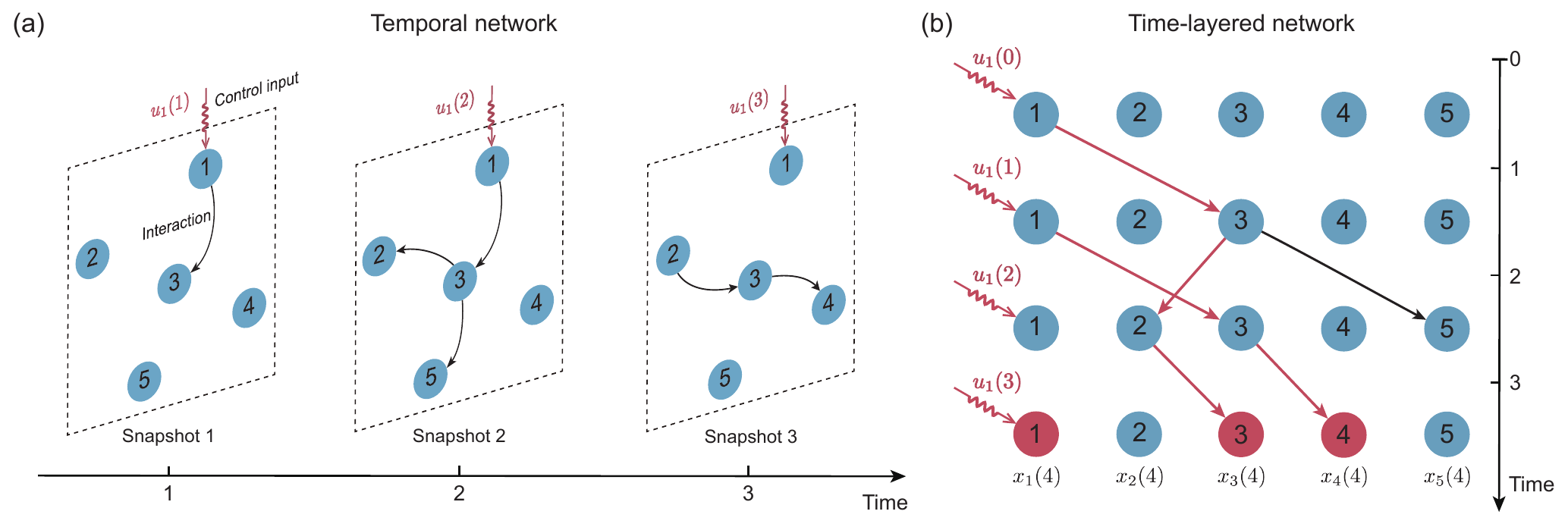}
	\caption{Illustration on a simple temporal network and the corresponding time-layered network. (a) The snapshot representation of a temporal network with five nodes and three snapshots. Here, node $v_1$ is the driver node, which directly receives the external control input marked in red.
		In each snapshot, the black arrows depict the directed interactions between nodes (blue) at the corresponding moment.
		(b) The corresponding time-layered representation of the temporal network shown in (a) from $t_0 =0$ to $t_1 = 4$.
		The node-disjoint paths from the driver node $v_1$ to the nodes at target time $t_1=4$ are highlighted in red.
		According to the independent path theorem~\cite{posfai2014structural}, nodes $v_1$, $v_3$ and $v_4$ are structurally controllable at target time $t_1=4$. }
	\label{fig1}
\end{figure*}

Consider a directed temporal network with a set of nodes $V=\left\{v_1,v_2,\cdots,v_N\right\}$.
Here we regard the temporal network as a collection of subnetworks that are chronologically ordered.
We refer to such subnetworks as snapshots, as shown in Fig.~\ref{fig1}(a).
In this paper, we consider the discrete time-varying linear dynamics
\begin{equation}
	\mathbf{x}(t+1) = \mathbf{A}(t)\mathbf{x}(t)+\mathbf{B}(t)\mathbf{u}(t),
	\label{beq1}
\end{equation}
where $\mathbf{x}(t) = \left[x_1(t),x_2(t),\cdots,x_N(t)\right]^{\mathrm{T}}\in \mathbb{R}^{N}$ represents the state of the system with $x_i(t)$ being the state of node $v_i$ at time $t$; $\mathbf{A}(t)\in \mathbb{R}^{N\times N}$ is the adjacency matrix at time $t$, indicating the interactions between nodes; $\mathbf{u}(t)=\left[u_1(t),u_2(t),\cdots,u_{n_{\mathrm{I}}(t)}(t)\right]^{\mathrm{T}}\in \mathbb{R}^{n_{\mathrm{I}}(t)}$ is the input vector and  $n_{\mathrm{I}}(t)$ indicates the number of external input signals at time $t$. The matrix $\mathbf{B}(t)\in \mathbb{R}^{N \times  n_{\mathrm{I}}(t)}$ is the input matrix, which tells the injection locations of all external inputs. The nodes that directly receive external control inputs are called ``driver nodes".
Here, we assume all driver nodes are fixed throughout all snapshots,~i.e.,~$\mathbf{B}(t) = \mathbf{B}$.

The temporal network can also be represented as a time-layered network, as shown in Fig.~\ref{fig1}(b).
	For each node $v_i \in V$ at each time step $t >t_0$, we create a copy $\hat{v}_{(i, t)}$.
	If there exists a temporal link from node $v_i$ to node $v_j$ at time $t$, denoted by $\left(v_i, v_j, t\right)$, then we connect the nodes $\hat{v}_{(i, t)}$ to $\hat{v}_{(j, t+1)}$ in the time-layered network.
In a temporal network, a time-respecting path is a sequence of temporal links, where one edge at time $t$ ends at the starting node of another at $t+1$, for example $(v_i, v_j, t)$ is followed by $(v_j, v_k, t+1)$.
Two time-respecting paths are considered independent if they do not traverse the same node at the same time.
By introducing the time-layered representation, independent time-respecting paths in a temporal network are equivalent to node-disjoint paths in the corresponding time-layered network.


System~\eqref{beq1} is said to be controllable at target time $t _1$ in finite $\Delta t=t_1-t_0$ time steps, if it can be driven from any initial state $\mathbf{x}(t_0)$ to any desired final state by appropriately setting $\mathbf{u}(t)$.
Due to the property of linearity, we can shift $\mathbf{x}(t)$ so that the initial state $\mathbf{x}(t_0)=\mathbf{0}$.
Then we can write the system state at time $t_1$ as
\begin{equation*}
	\begin{aligned}
		\mathbf{x}\left(t_{1}\right)= &\mathbf{A}\left(t_{1}-1\right) \mathbf{A}\left(t_{1}-2\right) \cdots \mathbf{A}\left(t_{0}+1\right) \mathbf{B}\left(t_{0}\right) \mathbf{u}\left(t_{0}\right)\\
		& +\cdots+\mathbf{A}\left(t_{1}-1\right) \mathbf{B}\left(t_{1}-2\right) \mathbf{u}\left(t_{1}-2\right)\\
		&+\mathbf{B}\left(t_{1}-1\right) \mathbf{u}\left(t_{1}-1\right).
	\end{aligned}
\end{equation*}
We set $ \mathbf{\Phi}(l,h) = \mathbf{A}\left(l-1\right)\mathbf{A}\left(l-2\right)\cdots \mathbf{A}\left(h\right)$ with $l>h$.
In particular, $\mathbf{\Phi}(t,t) =\mathbf{I}, \forall t\in\mathbb{N}$, where $\mathbf{I}$ is the identity matrix.
Then $\mathbf{x}\left(t_{1}\right)$ can be rewritten as
$$
	\mathbf{x}\left(t_{1}\right)= \sum_{t=t_0}^{t_1-1} \mathbf{\Phi}(t_1,t+1)\mathbf{B}(t)\mathbf{u}(t).
$$
By defining the temporal controllability matrix as
\begin{equation*}
\begin{aligned}
	\mathbf{C}(t_0,t_1)=&\left[\mathbf{\Phi}(t_1,t_0+1)\mathbf{B}(t_0),\cdots,\right. \\
	&\left. \mathbf{\Phi}(t_1,t_1-1)\mathbf{B}(t_1-2),\mathbf{B}(t_1-1) \right],
\end{aligned}
\end{equation*}
we get
$$
\mathbf{x}\left(t_{1}\right)=\mathbf{C}(t_0,t_1)\mathbf{U},
$$
where $\mathbf{U} =\left[\mathbf{u}^{\mathrm{T}}(t_0),\mathbf{u}^{\mathrm{T}}(t_0+1),\cdots,\mathbf{u}^{\mathrm{T}}(t_1-1)\right]^{\mathrm{T}}$.
It is clear that $(\mathbf{A}(t), \mathbf{B}(t))$ is controllable, if
\begin{equation}
	\operatorname{rank}(\mathbf{C}(t_0,t_1)) = N.
	\label{condition}
\end{equation}

However, in actual scenarios, the precise measurement of edge weights is often infeasible. Moreover, due to complexity for large-scale temporal networks, determining the rank of $\mathbf{C}(t_0,t_1)$ incurs prohibitively high computational cost.
Therefore, we concentrate on investigating the structural controllability of temporal networks.
The system $(\mathbf{A}(t),\mathbf{B}(t))$ is structurally controllable at target time $t_1$ in $\Delta t = t_1-t_0$ time steps, if there exist choices of non-zero weights in $(\mathbf{A}(t),\mathbf{B}(t))$ such that the system satisfies Eq.~\eqref{condition}.
According to the independent path theorem~\cite{posfai2014structural}, the subset $C\subseteq V$ is a structurally controllable subset at target time $t_1$ in $\Delta t$ time steps, if and only if all nodes $v_i \in C$ at time $t_1$ are connected from driver nodes within $\left(t_1-\Delta t,t_1\right]$ through independent time-respecting paths.
Thus, the temporal network is structurally controllable at target time $t_1$ in $\Delta t$ time steps, if and only if $C=V$ holds.
Furthermore, given a set of driver nodes, we can determine the dimension of the maximum controllable subspace,~i.e.,~the cardinality of the maximum structurally controllable subset.
This can be achieved by detecting the max-flow in the time-layered network from the driver nodes within $\left(t_1-\Delta t,t_1\right]$ to all nodes $v_i \in V$ at time $t_1$ with the capacity of each edge and each node being 1.
To impose restrictions on nodes, we can split each node into the ``in" and ``out" nodes connected by an edge with the capacity of 1.
Then we connect all the incoming edges of the original node to the new ``in" node, and all the outgoing edges of the original node to the new ``out" node.
This modification allows us to obtain an auxiliary graph accordingly and further apply the standard max-flow algorithm~\cite{ravandi2019identifying}.

\section{Online Time-accelerated Heuristic Algorithm}


For a given temporal network, we focus on finding the minimum set of driver nodes to guarantee network controllability.
However, it remains an NP problem in terms of time complexity, so we aim to identify a set of driver nodes, which is a suboptimal solution.
While there may exist more than one maximum controllable subspaces for a set of driver nodes, their dimensions are identical.
Based on that, it is straightforward that a set of driver nodes and the dimension of its corresponding maximum controllable subspace satisfy the condition of mapping.

We map a set of driver nodes $D$ to the dimension of its corresponding maximum controllable subspace $N_{\mathrm{c}}$ by a set function
\begin{equation}
	f(D)=N_{\mathrm{c}}, \ D\in 2^{V},
	\label{beq_set}
\end{equation}
where $2^{V}$ denotes the power set of $V$,~i.e.,~the set of all subsets of $V$, including the empty set and $V$ itself.
It is clear that the set function is monotonically non-decreasing, that is, for all subsets $P\subseteq V$, the set function satisfies
\begin{equation*}
	f(P \cup\{v_i\})-f(P) \geqslant 0, \ \forall  \ v_i \in V \backslash P,
\end{equation*}
where $V \backslash P=\left\{x | x\in V, x\notin P\right\}$ represents the set difference between $V$ and $P$.

Identifying the minimum number of driver nodes to fully control temporal networks can be formulated as
\begin{equation*}
		\begin{aligned}
			\min\limits_{D \subseteq V} ~&\vert D \vert\\
			\operatorname{s.t.}~&f(D)=N.
		\end{aligned}
\end{equation*}
Namely, the goal is to identify as few driver nodes as possible ($\vert D \vert$) while ensuring that the dimension of its corresponding maximum controllable subspace equals $N$.
A widely used  approach to the combinatorial optimization problem is the greedy algorithm. 
For $v \in V$, $D\in 2^{V}$, let $\Delta(v|D)  = f(D\cup \left\{ v \right\})- f(D)$ denote the discrete derivative of $f$ at $D$ with respect to $v$, also called the marginal gain of the element $v$ with respect to the set $D$. Let $D_k$ denote the set returned by the $k$th iteration.
Starting with $D_0 = \emptyset$ and in iteration $k$, we have
\begin{equation*}
	D_{k} = D_{k-1}\cup \left\{ \mathop{\arg\max}\limits_{v}\Delta(v| D_{k-1})\right\}.
\end{equation*}

\begin{figure}[htbp]
	\centering
	\includegraphics[width=0.475\textwidth]{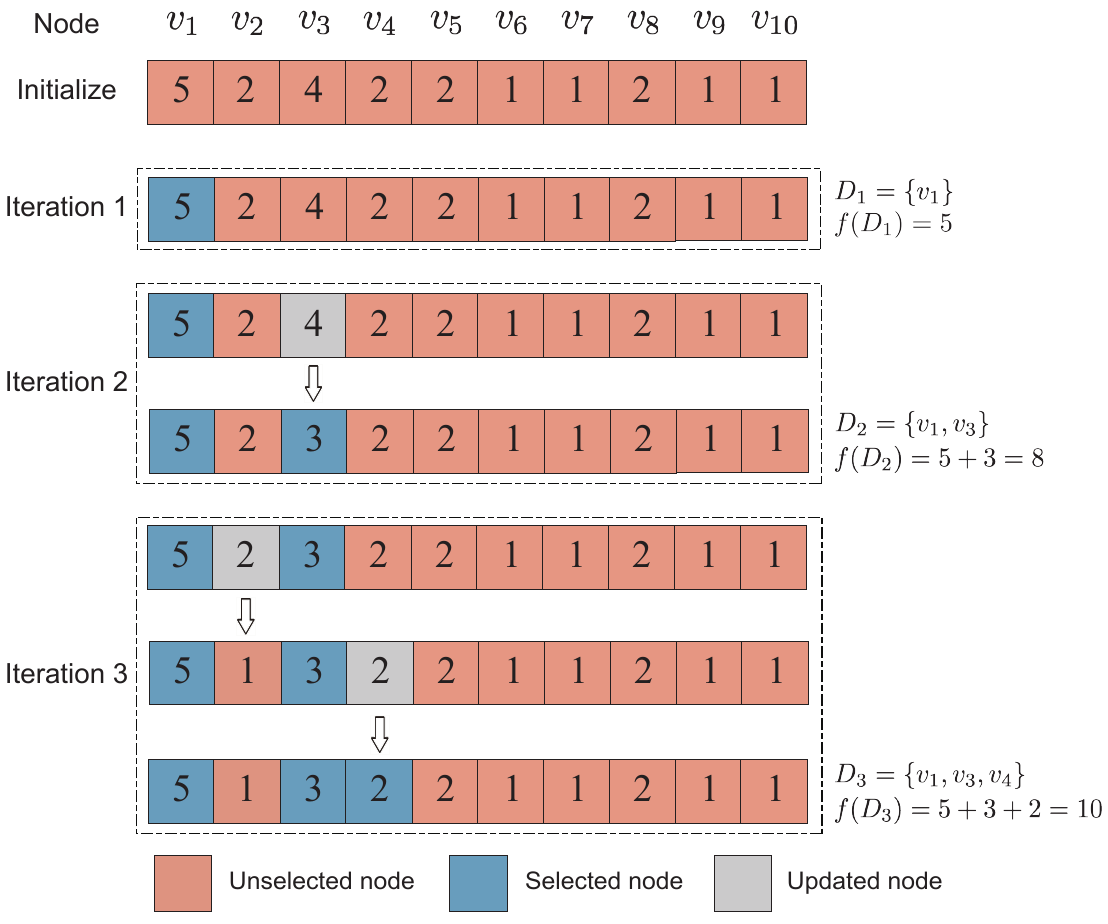}
	\caption{
		Illustration on the process of selecting driver nodes.
		Here, we take an example of a tempotal network with 10 nodes.
		Each box represents a node and the number represents the marginal gain of the node,~i.e.,~the increment of the dimension of the maximum controllable subspace when adding the node as a driver node.
		First, we initialize the increment array by calculating the  dimension of the maximum controllable subspace of each node independently (shaded in pink).
		Then in iteration 1, we select node $v_1$ corresponding to the maximum value, which leads to the corresponding set of driver nodes $D_1=\left\{v_1\right\}$ (blue) and controllable space $f(D_1)=5$.
		In subsequent iterations, we select a new driver node from the unselected nodes (pink).
		During the $i$th iteration ($i>1$), unlike the greedy algorithm which updates all the unselected nodes simultaneously, we only update the node (grey) that has the maximum marginal gain among unselected nodes.
		If the updated marginal gain of one node remains the maximum among unselected nodes, it will be selected (blue).
		In iteration 2, we update node $v_3$, which has the maximum value 4 (grey) among unselected nodes, and find that its marginal gain 3 (blue) remains the maximum, resulting in $D_2=\left\{v_1, v_3\right\}$ and $f(D_2)=8$.
		In iteration 3, we first update node $v_2$, and find that the marginal gain of node $v_2$ is 1, which is not maximum among unselected nodes.
		Thus, node $v_2$ is not selected and still marked in pink.
		We then update node $v_4$, and find that its marginal gain remains maximum, yielding $D_3=\left\{v_1, v_3,v_4\right\}$ and $f(D_3)=10$.
		As $f(D_3)=10$ equals to the number of nodes, the iteration ends.
		This process reduces the number of traditional calculations required by the greedy algorithm from 27 to 13.}
	\label{increment}
\end{figure}

Further, we introduce the submodularity of a set function.
For a finite set $V$, a set function $g: 2^{V} \rightarrow \mathbb{R}$ is submodular, if it satisfies one of the following equivalent conditions \cite{schrijver2003combinatorial}:
\begin{itemize}
	\item For any $P,Q \subseteq V$, $g(P)+g(Q) \geqslant g(P \cup Q)+g(P \cap Q)$ holds;
	\item For any $P\subseteq Q \subseteq V$ and any $x\in V\backslash Q$, $g(P\cup \left\{x\right\})-g(P)\geqslant g(Q\cup \left\{x\right\})-g(Q)$ holds;
	\item For any $P \subseteq V$, and $x_1, x_2 \in V\backslash P$ such that $x_1 \neq x_2$, $g(P\cup \left\{x_1\right\})+ g(P\cup \left\{x_2\right\}) \geqslant g(P\cup \left\{x_1,x_2\right\}) +g(P)$ holds.
\end{itemize}

By employing the property of the cut between two disjoint node sets in a graph~\cite{ford2010flows}, the  submodularity of the set function \eqref{beq_set} can be obtained~\cite{srighakollapu2021optimizing}.
Here we rely on the property of submodularity to employ pruning strategies to eliminate subsets of the search space that are unlikely to yield the maximum increment. This approach can converge more quickly to the desired solution, as illustrated in Fig.~\ref{increment}.
This accelerated version is a practically useful method for the fast identification of driver nodes in temporal networks~\cite{minoux2005accelerated}.

Furthermore, the submodularity of the set function $f$ can not only be used for time acceleration but also give an upper bound estimation for the solution accuracy~\cite{wolsey1982analysis}. Let $D_0= \emptyset, D_1,\cdots,D_k,\cdots$ be the sequence of sets obtained by the greedy iteration, where the subscript $k$ indicates the number of elements in the set.
For the submodular set function $f$, we have
	\begin{equation*}
		\ell \leq\left(1+\ln f(D_1)\right) N_{\mathrm{D}},
	\end{equation*}
	where $N_{\mathrm{D}}$ denotes the minimum number of driver nodes and $\ell$ is the smallest index satisfying $f(D_{\ell})=N$ in the sequence of sets,~i.e.,~the number of driver nodes.

\begin{figure*}[htbp]
	\centering
	\includegraphics[width=.95\textwidth]{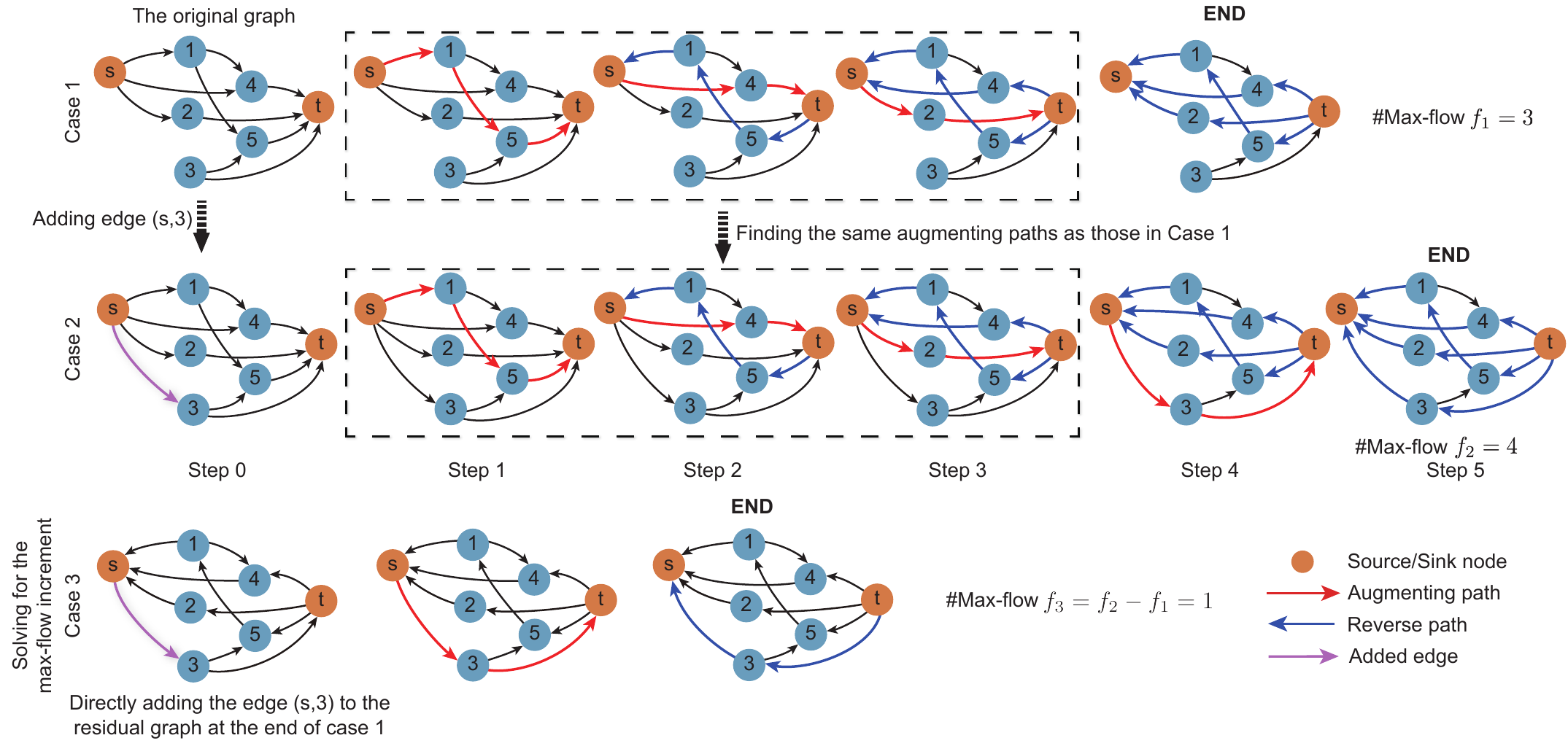}
		\caption{Solving for the max-flow increment by the modified residual graph. 
		The augmenting paths, paths from the source node  to the sink node (marked in orange), are marked with red edges.
		In each step of case $i$, we detect an augmenting path, replace it with the corresponding reverse path (blue), and obtain a residual graph.
		In case 1, after three steps, we find that there is no augmenting path, and derive that the max-flow on the original graph is three, i.e., $f_1 = 3$.
		In case 2, we add edge $(\mathrm{s},3)$ (purple) to the original graph, and in the first three steps, we assign the same augmenting paths as those (boxed with dashed lines) in case 1.
		In the fourth step, we continue searching for the augmenting path and find the max-flow $f_2 = 4$.
		In case 3, we add the edge $(\mathrm{s},3)$ to the residual graph at the end of case 1, and find that the max-flow $f_3=1$ is equal to the increased max-flow obtained by adding edge $(\mathrm{s},3)$ to the original graph (case 2),~i.e.,~$f_3 = f_2-f_1=1$).
		Thus, to solve for the max-flow increment, we can employ the steps outlined in case 3, which effectively eliminates three unnecessary steps in comparison to case 2.
	}
	\label{fig_residual}
\end{figure*} 
As aforementioned, it is crucial to determine the increment of the dimension of the maximum controllable subspace when adding a  new node into the current driver node set.
Note that given a set of driver nodes, we can determine the maximum controllable subspace by identifying the corresponding maximum independent time-respecting paths.
Specifically, this problem can be addressed by determining the max-flow in the auxiliary graph, which involves connecting the source node to the ``in" driver nodes across all time layers and linking all the ``out" nodes at the target time to the sink node.
We typically employ the Ford-Fulkerson algorithm for solving the maximum flow problem, which involves initializing the residual graph $\mathcal{R}$ to the auxiliary graph and then updating it dynamically.
Specifically, the Ford-Fulkerson algorithm iterates the following two steps:
\begin{itemize}
	\item[(1)] Search for an augmenting path in the residual graph $\mathcal{R}$. Here an augmenting path is a path from the source node  to the sink node.
	If there is no augmenting path, the iteration procedure ends.
	\item[(2)] Delete the augmenting path and add the reverse unit-weighted augmenting path with the max-flow plus 1.
\end{itemize}

As long as there is an augmenting path in the residual graph $\mathcal{R}$, the found path will increase the total flow; otherwise, the total flow has achieved its maximum feasible value and then the algorithm ends.
Generally, to obtain the marginal gain $\Delta(v_i|D_{k-1})$, we need to compute the max-flow on the graph for the set of driver nodes $D_{k-1}\cup \left\{v_i\right\}$.
As the algorithm iterates, the cardinality of the set $D_{k-1}$ increases, which leads to an increased computational cost for calculating the set function $f(D_{k-1}\cup \left\{v_i\right\})$.
Consequently, the computational cost associated with updating the marginal gains of nodes also escalates, resulting in increased difficulty for executing the algorithm.

In the Ford-Fulkerson algorithm, the reverse flow operation, often known as the backflow operation, plays a vital role.
By allowing flow to be pushed back along edges in the opposite direction, the algorithm can effectively ``undo" the previous flow assignment and redistribute the flow through a more effective path.
Consequently, the order of choosing augmentation paths does not matter for the outcome.
In the max-flow algorithm, adding a new driver node corresponds to adding new edges from the source to the ``in" driver nodes throughout all time layers.
It is emphasized that we focus on the increment of the max-flow by adding new edges, rather than the absolute value of the max-flow itself.
Figure~\ref{fig_residual} illustrates the operation to obtain the max-flow increment by adding edges to the residual graph at the end of last iteration.
Therefore, we can treat the marginal gain of a node on each iteration as the max-flow on the residual graph with added edges. 
The approach involves online learning of the max-flow in a dynamic graph with edges added incrementally.
The concept of online learning is originally introduced in the study of data stream processing.
Here, we adopt a variant of this methodology, which eliminates unnecessary steps by directly utilizing results from the previous iteration to determine the max-flow increment, and further reduce computational cost.

Considering the time-varying network structure represented by $\mathbf{A}(t)$, we propose an online time-accelerated heuristic algorithm (OTaHa) to identify a set of driver nodes in temporal networks.
The pseudo code for the online time-accelerated heuristic algorithm is provided in Algorithm~\ref{algo1}, and the pseudo code for determining the max-flow increment, a critical component of our proposed algorithm, is outlined in Algorithm 2.

\begin{algorithm}[htbp]
	\SetAlgoLined
	
	\SetKwRepeat{Do}{do}{while}
	\KwIn{The network structure represented by $\mathbf{A}(t)$, the number of nodes $N$ and the time steps $\Delta t$} 
	\KwOut{the set of driver nodes $D^{k+1}$} 
	\textbf{Initialize:} Let $ D^{0} = \emptyset, k=0$. Creat a residual graph list $\mathcal{R}$. Construct an auxiliary graph $\mathcal{G}_{\mathrm{a}}$ with the source node $\mathrm{s}$ connecting no nodes and the sink node $\mathrm{t}$ connected from all the ``out" nodes at the target time. 
	
	\For{$ i \leftarrow 1$ \KwTo $N$}{
		$\left[\Delta (v_i), \mathcal{R}(v_i)\right]$$\leftarrow$Online\_Maxflow($\mathcal{G}_{\mathrm{a}}, v_i, N, \Delta t$);
	}
	
	$\hat{v}\leftarrow \mathop{\arg\max}\limits_{v_i\in V} \left\{ \Delta(v_i) \right\}$;
	
	$\ D^{k+1}\leftarrow  D^{k}+\left\{\hat{v}\right\}$;
	
	\eIf{$f(D^{k+1})=N$}{
		\textbf{Return: $D^{k+1}$};
	}
	{
		\For{$k \leftarrow 1 $ \KwTo $N-1$}{

			\Do{$ \delta < \mathop{\max}\limits_{v_i\in V-D^{k},v_i \neq v_{i_o}} \left\{ \Delta(v_i) \right\}$ }
			{
				$v_{i_o} \leftarrow \mathop{\arg\max}\limits_{v_i\in V-D^{k}}\left\{\Delta(v_i)\right\}$;
				
				\If{$v_{i_o}$ has been picked once in the $k$th iteration}{
					
					$\delta \leftarrow  \Delta(v_{i_o})$;
					
					\textbf{break}\;
				}
				\textbf{	Calculate: }
				
				$\left[\delta, \mathcal{R}(v_{i_o})\right]\leftarrow$Online\_Maxflow($\mathcal{R}(\hat{v}), v_{i_o}, N, \Delta t$);
				
				$ \Delta(v_{i_o}) \leftarrow \delta $;
			}
			
			$D^{k+1}\leftarrow  D^{k}+\left\{ v_{i_o}\right\}$;
			
			$\Delta(v_{i_o})\leftarrow 0$;
			
			$\hat{v} \leftarrow v_{i_o}$;

			\If{$f(D^{k+1})=N$}{
				\textbf{Return: $D^{k+1}$}\;
				
			}
			
		}
	}
	\caption{The online time-accelerated heuristic algorithm}
	\label{algo1}	
	
\end{algorithm}

Lines 11-20 in Algorithm 1 constitute the core code for selectively updating the marginal gains of nodes.
Algorithm 2  provides the function ``Online\_Maxflow", which carries out the solution process for the max-flow increment, as shown in Fig.~\ref{fig_residual}.

\begin{algorithm}
	\caption{Online\_Maxflow}
	\SetKwRepeat{Do}{do}{while}
	\KwIn{The residual graph $\mathcal{R}$, a new driver node $v_i$, the number of nodes $N$ and the time steps $\Delta t$} 
	\KwOut{The max-flow increment $\Delta (v_i)$, the residual graph $\mathcal{R}(v_i)$} 
	\textbf{Initialize:} $\mathcal{R}(v_i) \leftarrow \mathcal{R}$, $\mathcal{P}(v_i) = \emptyset, \Delta (v_i) = 0$. 
	
	Connect the source node $\mathrm{s}$ to the ``in" driver nodes $v_i$ in all  time layers in $\mathcal{R}(v_i)$\;
	$\left[\Delta (v_i), \mathcal{R}(v_i)\right] \leftarrow$ Maxflow($\mathcal{R}(v_i),\mathrm{s},\mathrm{t}$)\;
	\textbf{Return:} $ \Delta (v_i), \mathcal{R}(v_i)$
	\label{algo22}
\end{algorithm}

	While our proposed algorithm (OTaHa) is limited to finding a locally optimal set of driver nodes, it comes with the potential for further methodological improvements in searching more solutions.
	One way is to initialize the set of driver nodes by randomly or strategically selecting some nodes.
	By injecting these slack variables into OTaHa, we can obtain multiple driver node sets.


To validate the performance of OTaHa, we compare its computational complexity with that of existing algorithms.
Denote the number of edges in the auxiliary graph as $E=N\Delta t+M$ (if we assume that the system state can be retained,~i.e.,~ there are self-loops for each node in all snapshots, $E=2N\Delta t+M$), where $M$ is the number of total edges in all snapshots, and $\Delta t = t_1-t_0$ is the number of time layers. The time complexity of the Ford-Fulkerson is $O(EF)$, where $F$ is the value of maximum flow. In the worst case, since each iteration is based on the residual graph of the last iteration and the temporal network has $N$ nodes,  we need to execute the max-flow algorithm at most $N$ times, where $F = N$. Therefore, the computational complexity of our proposed algorithm is $O(N^2 E)$,~i.e.,~$O(N^3 \Delta t+N^2M)$. The computational complexity of the heuristic algorithm and the greedy algorithm is $O(N^4+N^3 \Delta t+N^2M)$ and $O(N^4 \Delta t+N^3M)$, respectively.
Hence, the OTaHa significantly reduces the time complexity with respect to existing algorithms.

\section{Performance of OTaHa\label{algorithm}}

To verify the effectiveness of OTaHa, we next test the performance on some real and synthetic datasets.
We construct temporal networks from the following empirical datasets and basic statistical information is shown in Table~\ref{tab_basic}.
\begin{table*}[htbp]
	\renewcommand\arraystretch{1}
	\caption{Temporal characteristics of the empirical datasets. $N$ is the number of nodes, and $M$ is the number of edges. ``Num. of snapshots" represents the number of non-blank snapshots. Time resolution corresponds to the time window we set to construct each snapshot.}
	\centering
	\begin{ruledtabular}
		\begin{tabular}{c|cccccc}
			Dataset & $N$&  $M$   &Num. of snapshots & Time resolution  & Start time &End time\\
			\hline
			Colony 1-1 & 89 & 1,911 & 883 & 1s &0s & 1,438s \\ 
			Colony 1-2 &72 & 1,820&1,048 & 1s & 0s & 1,749s   \\ 
			Colony 2-1 &71&  975& 690& 1s & 0s&1,438s \\
			Colony 2-2 & 69 & 1,917 & 1,112 & 1s& 1s & 1,796s \\ 
			Colony 6-1 &33 & 652& 537 &1s &1s& 1,918s  \\ 
			Colony 6-2 & 32 & 367 & 312 & 1s & 2s& 1,755s  \\ 
			Enron email &89 &5,219 &253& 1d &1999-06-25 &2000-11-27  \\ 
			High school contacts & 126 & 28,561& 5,609& 20s & 54,120s &326,450s  \\ 
			Infectious contacts & 200 & 5,943 & 1,238 & 20s & 1,240,913,019s& 1,240,941,099s \\ 
			DPPI network &143 & 1,959 & 36 & -& -&-  \\ 
		\end{tabular}
	\end{ruledtabular}
	\label{tab_basic}
\end{table*}

\begin{table*}[htbp]
	\renewcommand\arraystretch{1}
	\caption{Performance comparison of existing algorithms and our algorithm (OTaHa) for detecting driver nodes in the ant interaction dataset  indicated by Colony ID.
		$N_{\mathrm{D}}$ is the minimum number of driver nodes in the dataset obtained by using the brute force algorithm.
		The number of optimal solutions is represented by ``Num.~of sets".
		The smallest numbers of driver nodes obtained by the heuristic algorithm, the greedy algorithm and the proposed algorithm are denoted by $N_{\mathrm{h}}$, $N_{\mathrm{g}}$ and $N_{\mathrm{OTaHa}}$, respectively.
		We use ``bold" to indicate the minimum execution time. }
	\centering
	\begin{ruledtabular}
		\begin{tabular}{C{2cm}|C{1cm}C{2.5cm}|C{1cm}C{2.5cm}|C{1cm}C{2.5cm}|C{1cm}C{2.5cm} }
			\multirow[b]{2}{*}{\makecell[c]{Colony ID}} & \multicolumn{2}{c|}{Brute force} 
			& \multicolumn{2}{c|}{Heuristic algorithm}&\multicolumn{2}{c|}{ \makecell{Greedy algorithm}} &\multicolumn{2}{c}{ \makecell{OTaHa}}\\ \cline{2-9}
			&$N_{\mathrm{D}}$&Num.~of sets& $N_{\mathrm{h}}$ &\makecell[c]{Execution time\\(seconds)} & $N_{\mathrm{g}}$ &\makecell[c]{Execution time\\(seconds)}  & $N_{\mathrm{OTaHa}}$ &\makecell[c]{Execution time\\(seconds)}\\ 
			\hline
			1-1 & \cellcolor{Lavender}3 & 153 & 7 & 216.55 & 3  & 83.28&\cellcolor{Lavender}3  & \textbf{5.72}\\ 
			1-2 & \cellcolor{Lavender}2  & 21& 7 & 134.05 &2 & 40.89&\cellcolor{Lavender}2  &\textbf{3.32}\\ 
			2-1 & \cellcolor{Lavender}5 &  89&  9 & 67.17 & 5 &73.23&\cellcolor{Lavender}5  & \textbf{3.33}    \\
			2-2 & \cellcolor{Lavender}4  & 115 & 4 & 55.17 &4 & 112.47&\cellcolor{Lavender}4& \textbf{5.24}  \\ 
			6-1 & \cellcolor{Lavender}1 & 3 & 1 &3.55& 1 & 1.20&\cellcolor{Lavender}1  & \textbf{0.31}\\ 
			6-2 & \cellcolor{Lavender}2 & 10 & 4 & 3.04& 2 & 1.46&\cellcolor{Lavender}2 &\textbf{0.36}\\ 
		\end{tabular}
	\end{ruledtabular}
	\label{tab_N_h}
\end{table*}

Here, we give some brief descriptions of these datasets.
Ant interactions~\cite{blonder2011time}: The ant interaction dataset captures all interactions between all ants.
Colonies were filmed in high definition with a digital camcorder.
In each film, each ant is recognized uniquely.
The interaction was defined as any part of one ant's (the initiator's) tentacles touching another ant's (the target's).
The colony interaction dataset contains six ant colonies.
Enron email dataset~\cite{michalski2011matching}: The dataset is a collection of emails (edges) exchanged by employees (nodes) of the Enron Corporation, released during the Federal Energy Regulatory Commission (FERC) investigation.
The dataset contains hundreds of thousands of emails, sent and received between 1998 and 2002.
High school dynamic contact networks~\cite{fournet2014contact}: This dataset provides the contacts (edges) between students (nodes) from three different classes in a high school located in Marseilles, France.
The dataset was collected over a period of 4 days in Dec. 2011.
Infectious SocioPatterns dynamic contact networks~\cite{isella2011s}: This dataset contains the daily dynamic contacts (edges) between the persons (nodes) collected during the Infectious SocioPatterns event that took place at the Science Gallery in Dublin, Ireland.
Dynamic protein-protein interaction (DPPI) network~\cite{DBLP:conf/bigdataconf/FuH22}: Here, a dynamic protein-protein interaction network is constructed using activity and co-expression analysis.
Specifically, this dataset stores temporal connections of the generated dynamic network, where a node represents a gene coding protein retrieved from Saccharomyces Genome Database, and an edge represents a protein-protein interaction in terms of protein functional associations at a certain timestamp.

In order to ensure state retention for all nodes, we assume that all nodes have self-loops in each snapshot.
It should be noted that analyses with and without state retention are permissible.
Next, we test the performance of these algorithms on the ant interaction dataset.
Detailed numerical results are represented in Table \ref{tab_N_h}, which show that the sets of driver nodes chosen by OTaHa are the optimal solutions for six ant colonies.
And these optimal values are approximately half of what the heuristic algorithm gets.
Additionally, we compare the corresponding execution time of these algorithms, where the execution time of OTaHa is significantly lower than the other two algorithms.
Among the six ant colonies, the execution time of OTaHa in colony ``2-1" exhibits the most significant reduction compared to the greedy algorithm. In this case, OTaHa achieves a remarkable 95.5\% decrease in execution time relative to the greedy algorithm, demonstrating its effectiveness and efficiency for identifying driver nodes.
Overall, in the ant interaction dataset, our proposed algorithm not only provides a high-quality collection of driver nodes, but also performs well with respect to execution time.

\begin{table}[htbp]
	\renewcommand\arraystretch{1.2}
	\centering
	\caption{
		Comparison of three algorithms on four empirical datasets.
		The term ``Num. of nodes" refers to the number of identified driver nodes.
		We use ``bold" to indicate the best performance.}
	\centering
	\begin{ruledtabular}
		\begin{tabularx}{.5\textwidth}{C{6em}|C{8.2em}C{3.6em}C{3.5em}c}
			
			Dataset &Metrics&Heuristic & Greedy &OTaHa \\
			\hline
			\multirow{2}{*}{Enron email} &Num.~of nodes &20&16&\textbf{16} \\
			& Execution time($\mathrm{s}$) &26.9  &161.0 & \textbf{1.8}\\   \cline{1-5}
			
			\multirow{2}{*}{\makecell[c]{High school\\contacts}}&Num.~of nodes & 15&8& \textbf{8}\\
			&Execution time($\mathrm{s}$) &6,955.7   &23,605.5 &\textbf{130.6} \\    \cline{1-5}
			
			\multirow{2}{*}{\makecell[c]{Infectious\\contacts}}&Num.~of nodes &66&45& \textbf{45}\\
			& Execution time($\mathrm{s}$)&2,739.1  & 34,556.0  &\textbf{55.4}\\  \cline{1-5}
			
			\multirow{2}{*}{DPPI}& Num.~of nodes &37&35& \textbf{35}\\
			& Execution time($\mathrm{s}$) &44.5   & 157.9  &\textbf{1.3}\\
			
		\end{tabularx}
	\end{ruledtabular}
	\label{other_dataset}
\end{table}
In addition, we make the analogous comparison of algorithm performances on the other datasets, and the specific results are shown in Table~\ref{other_dataset}.
Here, the optimal solution can not be derived from the brute force algorithm due to the large scale of these datasets.
It can be observed that our proposed algorithm outperforms existing algorithms on these datasets.
OTaHa consistently identifies fewer driver nodes than the heuristic algorithm.
In the high school dynamic contact dataset, for example, OTaHa detects approximately half the number of driver nodes compared to the heuristic algorithm.
For the Enron email dataset, the execution time decreases from 161.0 seconds to 1.8 seconds (a 98.9\% reduction); for the high school dynamic contact dataset, from 23,605.5 seconds to 130.6 seconds (a 99.4\% reduction); for the infectious SocioPatterns dynamic contact dataset, from 34,556.0 seconds to 55.4 seconds (a 99.8\% reduction); and for the DPPI dataset, from 157.9 seconds to 1.3 seconds (a 99.2\% reduction).
Generally, OTaHa requires less than two percent of the execution time compared to the greedy algorithm.
Furthermore, the larger the scale of the datasets we test, the more pronounced the advantage of OTaHa we observe for saving time.

\begin{figure}[htbp]
	\centering
	\includegraphics[width=.475\textwidth]{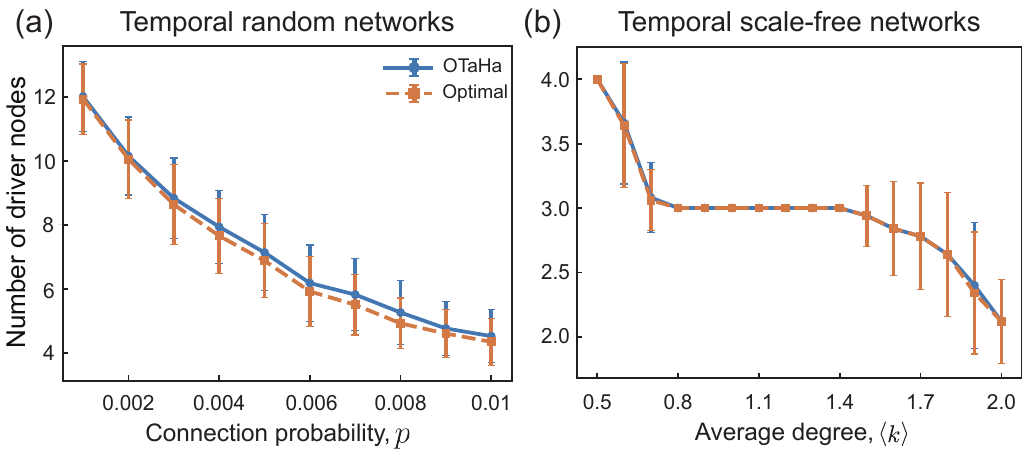}
	
	\caption{OTaHa detects the set of driver nodes that approximates the optimal solution on synthetic temporal networks.
		(a) Comparison with the optimal solution on temporal random networks with 15 nodes and 20 snapshots.
		The parameter $p$ denotes the probability of a node connecting to another under Erd\H{o}s-R\'enyi model for each snapshot~\cite{erdHos1960evolution}.
		(b) Comparison with the optimal solution on temporal scale-free networks with 40 nodes and 20 snapshots.
		The parameter $\langle k \rangle$ denotes the average degree of the scale-free network in each snapshot generated from the static model~\cite{goh2001universal}.
		Each data point represents the average number of driver nodes over 50 realizations.}
	\label{fig3}
\end{figure}

We also evaluate the effectiveness of our proposed algorithm on synthetic temporal random and scale-free networks generated by the Erd\H{o}s-R\'enyi model~\cite{erdHos1960evolution}~and the static model~\cite{goh2001universal}, respectively.
The results are depicted in Fig.~\ref{fig3}.
Apparently, OTaHa can identify a set of driver nodes in synthetic temporal networks, which closely approximates the optimal solution.

\section{Role of edges in the control of temporal networks}

Compared to static aggregated networks, the complete instantaneous signal-carrying paths can not be guaranteed in temporal networks where the interactions (edges) appears and disappears instantaneously.
This prompts us to study role of edges in the time-layered network for the control of temporal networks facing their possible failures or malicious attacks.
Indeed, edge attacks, which refer to the removal or disruption of some edges in a network, are frequently found in a broad variety of natural and engineered systems, such as connection failures in power grids, the damaging of critical roads in transportation networks, the inhibition of specific protein-protein interactions, and so on.
Such attacks can severely disrupt the normal function of a network and have a particularly significant impact on the network's controllability.
In certain networks, disabling just a single key edge may trigger a larger-scale disruption.
\begin{figure}
	\centering
	\includegraphics[width=.4\textwidth]{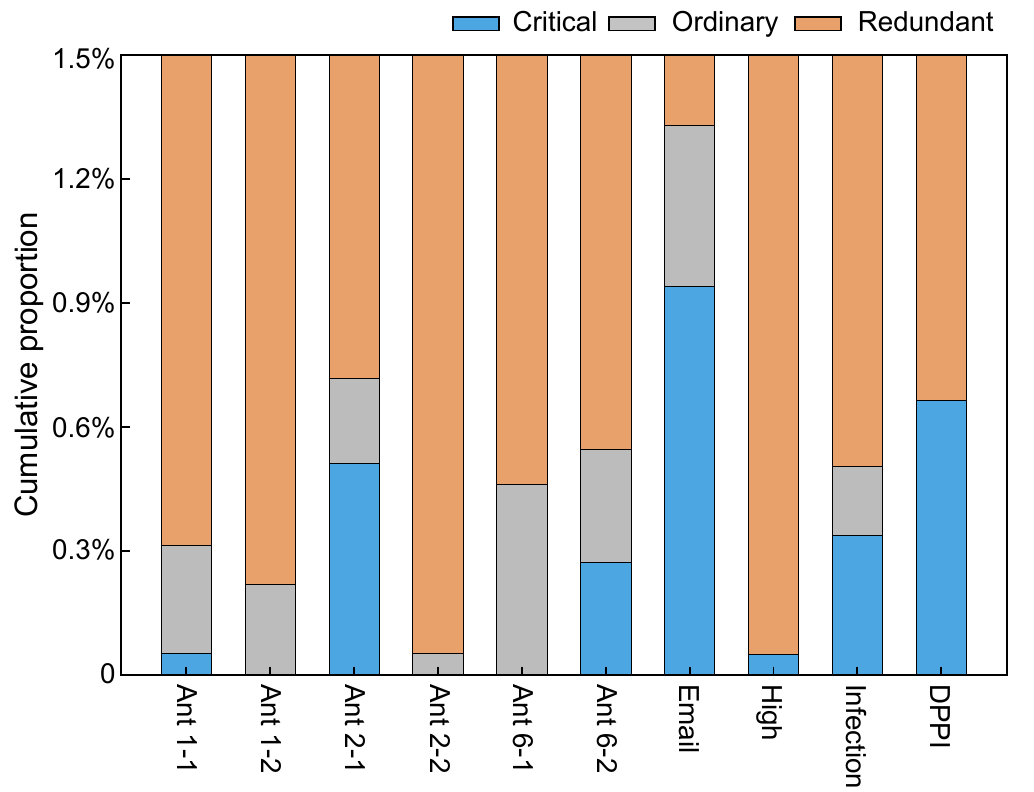}
	\caption{
		Classification of edges based on their role in controlling temporal networks.
		The horizontal axis represents different datasets, and the vertical axis indicates the fractions of critical (blue), ordinary (grey), and redundant (orange) edges.
		To better visualize and analyze the fractions of three edge types, we enlarge the 0-1.5\% range on the vertical axis containing critical and ordinary edges.
		We observe that there are only a small number of critical and ordinary edges.
	}
	\label{category}
\end{figure}

\begin{figure*}[htbp]
	\centering
	\includegraphics[width=\textwidth]{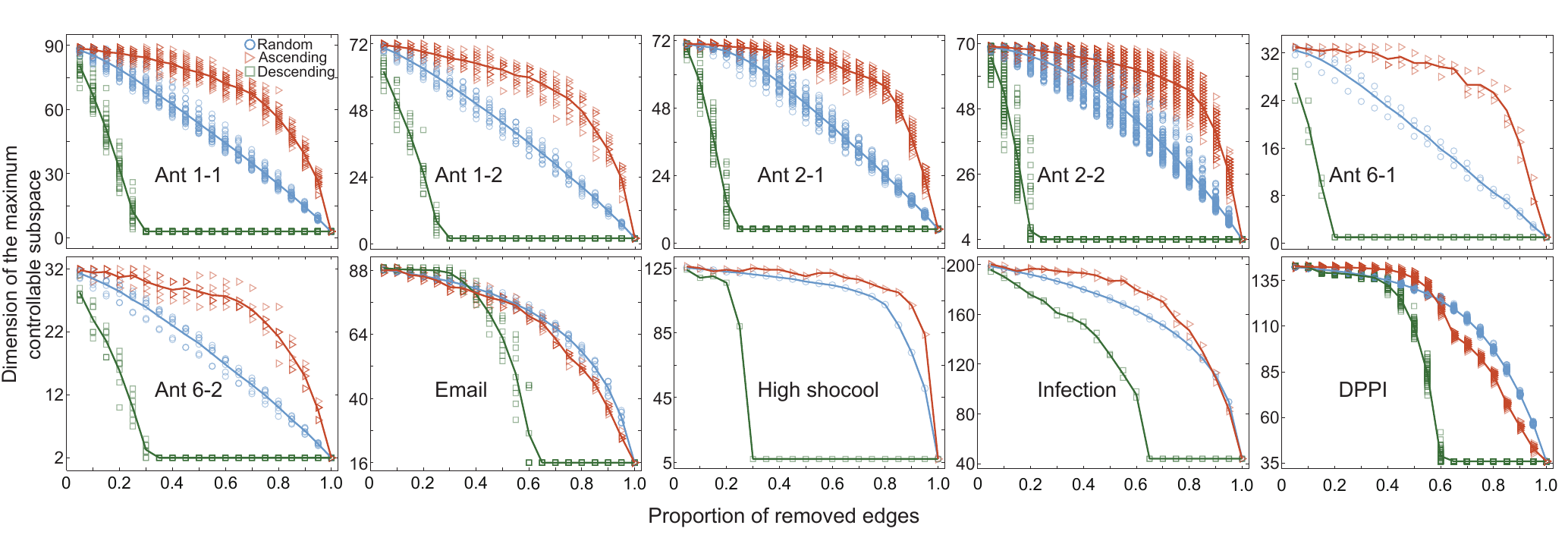}
	\caption{
		Removing edges with high edge betweenness dramatically impedes the controllability of temporal networks.
		We perform three edge attack strategies in removing network edges on different datasets following the random (circle), ascending (triangle), and descending (square) order of edge betweenness.
		The horizontal coordinate represents the proportion of removed edges, and the vertical coordinate represents the dimension of the maximum controllable subspace under different sets of driver nodes, which are represented by different markers in each panel.
		To ensure a comprehensive assessment of the network's robustness in the case of random edge attacks, the experiment is conducted 100 times.
		Self-edges indicate a state-preserving property and are not taken into consideration in the calculation of edge betweenness.}
	\label{edge_attack}
\end{figure*}

This observation leads us to classify the role of edges in a temporal network into three categories based on the effect of their failure on the control configuration: ``critical", ``redundant" and ``ordinary".
If a critical edge is removed, the number of driver nodes increases.
However, when a redundant edge is removed, the current set of driver nodes can still maintain full control of the network.
Moreover, if an ordinary edge is removed, it is possible to identify another set of driver nodes with the same number of driver nodes that maintain full control. This suggests that it may be sufficient to modify the configuration of the driver nodes without changing their number.
 In the analysis of the impact of edge removal on network controllability, as illustrated in Fig.~\ref{category}), we find that, in contrast to static networks, where most edges are classified as ordinary~\cite{liu2011controllability}, most edges in temporal networks are redundant.
  This suggests that, in temporal networks, there are few edges playing the critical role in determining the number of driver nodes.
  And the random failures or external attacks of some edges will not incur additional inputs required to maintain the full control of temporal networks.
  
Due to the interdependence of edges, the failure of one edge can trigger a chain reaction of subsequent failures.
In real-world networks, it is common for multiple edge failures to occur simultaneously or within a short time.
To further explore how the ability to control temporal networks evolves when some edges are removed or attacked, here we focus on the change of the dimension of the maximum controllable subspace for multiple sets of driver nodes.
Specifically, we investigate the effect of two typical edge attack (targeted and random) on the network controllability.
In targeted attacks, edges are intentionally selected based on their importance, while in random attacks, edges are removed without any particular pattern.
To assess the significance of edges forming the path carrying input signals to execute network control, we focus on the edge betweenness, which quantifies the number of shortest paths between any pair of nodes that pass through the edge.
We compute the edge betweenness in the time-layered network.
Then we carry out experiments under different attack strategies: random attacks, targeted attacks in descending order of edge betweenness, and targeted attacks in ascending order of edge betweenness.

The simulation results over five empirical datasets are presented in Fig.~\ref{edge_attack}.
The completely distinct performances of the two targeted attack strategies confirm the importance of edge betweenness in the robustness of achieving the controllability of temporal network.
In contrast to the random attacks, the dimension of the maximum controllable subspace declines dramatically when simply a fraction of edges are destroyed in the descending betweenness-based targeted attacks. 
The results suggest that when most of edges with high edge betweenness are removed, the network's controllability is substantially destroyed.

\section{Conclusions\label{conclusion}}

Recently, the concept of structural controllability has been extended to temporal networks and there is a growing body of literature focused on identifying driver nodes in temporal networks.
However, in large-scale temporal networks, existing algorithms typically exhibit inevitable limitations due to either low accuracy or high computational complexity.
In this paper, we propose an efficient algorithm (OTaHa) for identifying driver nodes with an upper bound $N_{\mathrm{OTaHa}} \leq\left(1+\ln f(D_1)\right) N_{\mathrm{D}}$ and computational complexity $O(N^3 \Delta t+N^2M)$.
In order to reduce computational cost, we take advantage of the submodularity property of set function and successfully eliminate unnecessary calculations.
Furthermore, to streamline the search process, we transform the incremental update at each iteration into the max-flow on the residual graph with edges added.

Our proposed algorithm is also flexible and can be used to obtain multiple sets of driver nodes by modifying the initial empty set configuration.
Through a series of experiments, we demonstrate the effectiveness of OTaHa in detecting driver nodes, as it consistently provides high-quality solutions while maintaining time efficiency.
Finally, we also find that edges with high edge betweenness considerably influence the controllability of temporal networks, which holds significant implications for devising strategies to maintain control in dynamical systems.
Thus, our findings provide an analytical framework offering driver nodes for controlling temporal networks, and significant insights into the driver nodes and key edges that intervene system behavior, which open the new avenue for achieving the ultimate control of temporal networks.




		\nocite{*}
\bibliography{bib_OTaHa}

	\end{document}